# AI in Action: Accelerating Progress Towards the Sustainable Development Goals

Brigitte Hoyer Gosselink, Kate Brandt, Marian Croak, Karen DeSalvo, Ben Gomes, Lila Ibrahim, Maggie Johnson, Yossi Matias, Ruth Porat, Kent Walker, James Manyika

May 30, 2024

Abstract

Advances in Artificial Intelligence (AI) are helping tackle a growing number of societal challenges, demonstrating technology's increasing capability to address complex issues, including those outlined in the United Nations (UN) Sustainable Development Goals (SDGs). Despite global efforts, 80% of SDG targets have deviated, stalled, or regressed, and only 15% are on track as of 2023, illustrating the urgency of accelerating efforts to meet the goals by 2030. This brief draws on Google's internal and collaborative research, technical work, and social impact initiatives to show AI's potential to accelerate action on the SDGs and make substantive progress to help address humanity's most pressing challenges. The paper highlights AI capabilities (including computer vision, generative AI, natural language processing, and multimodal AI) and showcases how AI is altering how we approach problem-solving across all 17 SDGs through use cases, with a spotlight on AI-powered innovation in health, education, and climate. We then offer insights on AI development and deployment to drive bold and responsible innovation, enhance impact, close the accessibility gap, and ensure that everyone, everywhere, can benefit from AI.

1. INTRODUCTION

The UN SDGs are regarded as one of the best representations of humanity's goals and challenges and are adopted by the UN's 193 member countries. A 2023 UN Special Edition Report highlighted a considerable gap in achieving the SDG targets: currently, the world is on track to achieve only 15% of the goals, and lagging behind on about 80% (see United Nations). The data show (see United Nations) that these challenges disproportionately impact low- and middle-income countries (LMICs) (see World Bank). Developments in Information and Communication Technologies (ICTs) accelerated progress against the Millennium Development Goals (MDGs), the predecessors of the SDGs (see United Nations). AI has the potential to do the same for implementing and advancing progress on the SDGs (see Vinuesa et al).

AI is a general-purpose technology with broad applicability and multiple uses across sectors. It helps machines learn, adapt, and perform tasks that have the potential to assist people, businesses, and organizations (see Ben-Ishai et al). Additionally, AI capabilities will continue to advance and create opportunities for innovation across domains — for example, AI techniques developed for image recognition in healthcare can also be applied to assess signs of disease in crops or moisture content in soil. The UN Department of Economic and Social Affairs highlights AI's potential to meaningfully enable 134 or 79% of the 169 SDG targets (see Vinuesa et al). Our collaborative research confirms this trend, identifying more than 600 existing AI use cases supporting the SDG targets within the social impact sector alone[1] — a 300% increase in the use of AI to address the SDGs since 2018 (see McKinsey).

a. Methodology

AI is demonstrably impacting a wide range of societal challenges and opportunities in profound ways. This brief examines AI's potential to accelerate progress on the SDGs, using evidence, insights, and case studies from Google's research partnerships, technical innovations, and social impact initiatives, as well as secondary data sources:

Research partnerships: We draw on Google's collaboration with world-leading research and consulting organizations including McKinsey & Company (see McKinsey), Boston Consulting Group (see Dannouni et al), and Ipsos (see Google), which have explored AI's potential for advancing the SDGs,[2] including mapping AI to SDG targets, accelerating climate action, and understanding global public attitudes towards AI.

Technical innovations: Google's research teams continue to advance AI research and solutions, driving technical innovations to address societal challenges (including AlphaFold, Flood Hub, and Data Commons). Much of this work is done in collaboration with partners in the field and insights from these ongoing projects have informed our analysis.

Social impact initiatives: Over the last five years, Google.org has supported more than 200 social impact organizations globally using responsible AI to advance their work through initiatives including the AI Impact Challenge (see Google) and AI for Global Goals (see Google), run in partnership with the Centre for Public Impact. These efforts and the work of the many social impact organizations informed our perspectives on how to advance progress of the SDGs using AI. Examples of these social impact organizations are showcased throughout this brief to demonstrate real-world applications with tangible impact.

---

[1] In 2023, through collaborative research, a library of use cases was developed. This library will continue to evolve and thus should not be treated as exhaustive.

[2] For example: In 2023, through collaborative research, we surveyed about 60 experts across 17 countries and interviewed over 50 leaders from nonprofits, tech companies, and foundations globally to gather insights on AI's potential for the SDGs, its responsible deployment, and the challenges to scaling.

Secondary data sources: We also leverage existing secondary data sources, including research publications and reports from international organizations (e.g., the UN) and AI-focused institutions (e.g., Stanford Institute for Human-Centered Artificial Intelligence).

b. Factors driving advancements in AI

In the last five years, there has been a surge in the exploration, development, and deployment of AI solutions across sectors, including the social impact sector (see [McKinsey](#)). From 2021 to 2022, patents granted worldwide to those developing AI capabilities, applications, and systems increased by 63% and this number has increased more than 30-fold since 2010. In 2023, 149 foundation models were released, more than double the amount released in 2022. Over 65% of these were open source, compared to 44% in 2022 and 33% in 2021 (see [Stanford University](#)).

These advancements in AI, among others, have been driven by several factors, including:

Technical breakthroughs in AI architecture, where landmark work in unsupervised learning (see [Le et al](#)), instruction, fine-tuning, Reinforcement Learning from Human Feedback (RLHF), the invention of Transformers (see [Vaswani et al](#)), and the development of a new generation of AI computing infrastructure have laid the foundations for the development of increasingly sophisticated AI systems.

Open access models (see [United Nations](#)) and low code/no-code solutions (see [Lou et al](#)) have helped to democratize AI adoption, allowing a broader spectrum of users, including governments and nonprofits, to leverage AI (see [USAID](#)).

Rapid growth in cost-effective cloud computing has accelerated the deployment of AI-driven solutions and will continue to contribute to further advancements in AI (see [Gill et al](#)).

Open data initiatives have enhanced the availability and quality of data online, fueling AI-driven tooling, insights, and decision-making across various domains (see [European Union](#)).

Improved connectivity and internet penetration (particularly in previously inaccessible regions) have facilitated AI advancements through real-time data collection and analysis. However, expanding this reach further remains a development challenge (see [Ritchie et al](#)).

The growth of remote sensing tools and associated technologies (e.g., sensors, mobile phones, and satellites) have supported data abundance and increased overall monitoring and data capabilities (see [Precedence Research](#)).

Heightened focus on digital transformation (see [McKinsey](#)), coupled with growing demand and supply of AI skills in the workforce (see [Green and Lamby](#)), has led to increased investment in AI and the adoption of AI capabilities within organizations. Although not evenly distributed across the world, the expanding AI workforce has helped bridge the talent gap, facilitating the expansion of AI capabilities and the development of AI tools and solutions (see [Green and Lamby](#)).

Collaborative ecosystems that bring together industry, governments, researchers, developers, and civil society, have helped us to collectively expand and share knowledge on AI, identify ways to mitigate emerging risks, and further develop and implement AI solutions (see [Richardson](#)).

Altogether, these factors have led to an increasingly powerful AI stack that is being deployed to provide high-impact solutions — tools, systems, and processes — to address our toughest challenges and bring about positive outcomes for people.

However, despite progress, AI advancements remain geographically skewed. AI development and deployment are concentrated in specific regions (particularly North America and Europe), while other regions of the world face barriers like lack of funding, compute costs, and talent shortages. Regional leadership from bodies like the United Nations and World Bank, policies to support infrastructure and innovation, and targeted initiatives, like Google's AI research center in Ghana, are crucial to ensuring democratized AI access and empowering local communities to develop and deploy AI-driven solutions to societal challenges, like the ones outlined in the SDGs (see [Nelson and Walcott-Bryant](#)).

c. Leading AI capabilities

AI encompasses a range of technical capabilities that are accelerating progress toward the SDGs. In this section, we highlight six key AI capabilities: (1) <u>Computer Vision</u>: Empowers machines to interpret visual data from images and videos, enabling tasks such as image recognition, object detection, and classification. (2) <u>Natural Language Processing</u> (NLP): Enables machines to understand and generate human languages, facilitating tasks like sentiment analysis and code generation. (3) <u>Speech and Audio Processing</u>: Allows analysis and synthesis of audio signals, enabling applications like speech recognition and voice-controlled interfaces, speech synthesis, and spoofing detection. (4) <u>Multimodal Foundational Models</u>: Integrates and interprets information from various sources, including text, images, and audio, to enhance understanding and generate contextually rich content. (5) <u>Reinforcement Learning</u> (RL): Enables agents to learn optimal decision-making strategies by interacting with an environment and receiving feedback. (6) <u>Robotics</u>: Encompasses the design, construction, and programming of physical machines capable of autonomously performing tasks.

AI capabilities are helping social impact organizations and leaders process vast amounts of data to better understand problems and target interventions, personalize information for frontline workers and communities, and deliver new modes of interaction. For instance, through our social impact partnership with ARMMAN we have seen AI-powered predictions facilitate targeted interventions for pregnant women in rural India most at risk of dropping out of health programs. [See Box 1.] Quill.org is using advances in NLP and Generative AI to provide immediate coaching and feedback to students on writing and reading comprehension. [See Box 2.] Computer Vision technology is helping Climate TRACE use satellite imagery to analyze emissions data. [See Box 3.] These examples show how various AI capabilities can accelerate progress toward a range of SDGs.

2. AI FOR THE SUSTAINABLE DEVELOPMENT GOALS

AI is driving progress towards all 17 SDGs. [See Table 1:Illustrative use cases of AI mapped to SDGs.] Experts have highlighted AI contributions toward three goals in particular that we explore more deeply: SDG 3 (Good Health and Well-Being), SDG 4 (Quality Education), and SDG 13 (Climate Action).[3] These experts attribute the increased potential in these areas to several factors, including: (1) <u>Organized and accessible datasets</u> that facilitate the development and deployment of AI solutions (e.g., administrative and academic data in education settings). (2) <u>A concentration of skilled AI professionals</u> in these domains. (3) <u>The suitability of challenges in these domains to AI-driven solutions</u>, especially those aided by pattern recognition, optimization, and prediction. (4) <u>Examples of scaled use cases</u>, where AI applications have demonstrated impact and opportunities for full scale potential could still be realized. (5) <u>The presence of profitable and/or more mature business models</u> that can attract investment and drive further development of AI solutions.

While these SDGs currently have been identified as high potential for AI, there remains significant opportunity to continue to invest in and accelerate AI application across all SDGs.

---

[3] In 2023, through collaborative research, we surveyed about 60 experts across 17 countries and interviewed over 50 leaders from nonprofits, tech companies, and foundations globally to gather insights on AI's potential for the SDGs, its responsible deployment, and the challenges to scaling.

Table 1: Illustrative use cases of AI mapped to SDGs

| SDG | How AI Could Help (not exhaustive) |
|---|---|
| SDG 1: No Poverty | <ul><li>Machine learning with satellite imagery and survey data can extract socioeconomic indicators to generate visualizations and predictions of poverty in areas lacking survey data (see [Google](#)).</li><li>Natural Language Processing can analyze financial data to provide personalized financial advice (AI chatbots providing financial counseling to underserved populations) (see [Mirestean et al](#)).</li><li>AI technologies can contribute to economic growth and job creation through automation, increased productivity, and innovation (see [Ben-Ishai et al](#)).</li></ul> |
| SDG 2: Zero Hunger | <ul><li>Image classification can monitor soil and crop health (predict pest invasions, real-time seed and soil analysis, and weed management) to optimize crop yields (see [Dhanya et al](#)).</li><li>Multimodal analysis can minimize food waste and optimize distribution to create smart supply chains (see [Onyeaka et al](#)).</li><li>Machine learning can be used with satellite imagery, localized food prices, and conflict data to predict and address severe acute childhood malnutrition and forecast food insecurity for early action (see [Google](#)).</li></ul> |
| SDG 3: Good Health and Well-Being | <ul><li>Medical image analysis can enhance accuracy in disease detection and diagnosis to decrease mortality rates ( see [Rana and Bhusan](#)).</li><li>Speech recognition technology can transcribe doctor-patient interactions and clinical notes to reduce time spent on documentation and allow healthcare professionals to focus more on patient care (see [Wang et al](#)).</li><li>Reinforcement learning can provide personalized treatment for chronic diseases to adapt interventions based on individual responses to improve health outcomes (see [Gandhi and Mishra](#)).</li><li>Robotic exoskeletons and assistive robots can provide support and assistance to elderly individuals and people with disabilities to improve accessibility (see [Qiu et al](#)).</li></ul> |
| SDG 4: Quality Education | <ul><li>AI-powered tools can help teachers develop lesson plans, personalize learning materials, and adapt learning dynamically based on real time feedback (see [OECD](#)).</li><li>Multimodal AI-powered educational software can adapt to different learning styles to make learning processes more inclusive and effective (see [Lee et al](#) ).</li><li>Natural language processing on conversations between educators and non-English-speaking parents, coupled with machine learning analytics on parent and student profiles, can construct a multilingual family engagement platform and deliver personalized resources to help parents digitally connect with teachers regardless of language (see [Google](#)).</li></ul> |
| SDG 5: Gender Equality | <ul><li>Machine learning can detect gender bias in hiring algorithms</li><li>Generative AI-powered chatbots can provide mental health support, offer personalized conversations and assistance to individuals in need (see [Denecke et al](#)).</li><li>Natural language processing methods, such as topic modeling, psycholinguistic feature modeling, and audio signal processing, can be applied to voice recordings and chat transcripts from crisis call hotlines for women to escalate calls with a high risk of intimate partner violence (see [Google](#)).</li></ul> |
| SDG 6: Clean Water and Sanitation | <ul><li>AI-optimized water infrastructure management can prevent leaks and optimize usage (see [Kamyab et al](#)).</li><li>Big data analysis and deep learning can predict droughts and floods for better water management (see [Sun and Scanlon](#)).</li><li>Machine learning analytics can estimate evapotranspiration by leveraging thermal satellite images, weather data, and agriculture data, to help farmers provide more precise amounts of water needed to irrigate crops (see [Google](#)).</li></ul> |

| SDG | How AI Could Help (not exhaustive) |
|---|---|
| SDG 7: Affordable and Clean Energy | <ul><li>Reinforcement learning can optimize energy distribution for smart grid management (see [Bousnina and Guerassimoff](#)).</li><li>AI, specifically ML and CV, can transform the electric grid into a more reliable and efficient system (see [Tapestry](#)).</li><li>Predictive analysis can develop a monitoring and maintenance framework for wind turbines (see [Maron et al](#)).</li></ul> |
| SDG 8: Decent Work and Economic Growth | <ul><li>Generative AI-powered chatbots can automate repetitive tasks and optimize data collection, freeing up human potential for higher-value work (see [McKinsey](#)).</li><li>AI can analyze data from sensors and cameras to identify potential safety hazards and prevent workplace accidents (see [Shah and Mishra](#)).</li><li>Machine learning analytics on publicly available skills and occupational data can map an individual's skill set captured through a guided assessment directly to relevant occupations (see [Google](#)).</li></ul> |
| SDG 9: Industry, Innovation, and Infrastructure | <ul><li>Computer Vision can detect defects in manufacturing processes to improve product quality (see [Bughin et al](#)).</li><li>Predictive analytics algorithms can analyze historical data and market trends to forecast demand for raw materials, components, and finished products (see [Seyedan and Mafakheri](#)).</li><li>Computer vision on Google Street View and LIDAR images can help residents assess defensible space and identify flammable vegetation around their homes (see [Google](#)).</li></ul> |
| SDG 10: Reduced Inequalities | <ul><li>Assistive audio-visual speech recognition can aid the hearing impaired (see [Kumar et al](#)).</li><li>Assistive technologies for individuals with visual disabilities can provide auditory descriptions of the surrounding environment (see [Jaboob et al](#)).</li><li>Machine learning to image and text data can develop a dictionary of norm-revealing phrases that can be used as an alternative credit-scoring mechanism to make consumer lending more accessible to low-income individuals (see [Google](#)).</li></ul> |
| SDG 11: Sustainable Cities | <ul><li>Machine learning models can simulate peripheral vision and improve driver safety (see [Harrington et al](#)).</li><li>Robots equipped with sensors and cameras deployed in disaster-stricken areas can assist in search and rescue operations and deliver essential supplies (see [Murphy et al](#)).</li><li>Reinforcement learning can optimize traffic flow and reduce congestion and pollution (see [Dilmegani](#)).</li><li>Machine learning analytics on incident dispatch data and other correlative data (weather, anomalous events, city demographics, etc.) can build a predictive model around emergency response times for first responders in urban areas (see [Google](#)).</li></ul> |
| SDG 12: Responsible Consumption and Production | <ul><li>AI-guided product design can help further sustainability and circularity (see [Ghoreishi and Happonen](#)).</li><li>Predictive analytics can match supply and demand, reducing overproduction (see [Seyedan and Mafakheri](#)).</li><li>Deep neural net and image-recognition technology on garbage and waste management images can help automate the classification and sorting of recyclable items at waste management facilities (see [Google](#)).</li></ul> |
| SDG 13: Climate Action | <ul><li>Reinforcement learning can optimize energy distribution and use (see [Bousnina et al](#)) traffic flow, and anti-pollution (see [Dilmegani](#)).</li><li>Machine learning analytics and computer vision on emissions data, satellite imagery of power plants, weather conditions, and grid conditions can monitor power plant emissions (see [Google](#)).</li></ul> |
| SDG 14: Life | <ul><li>Image and acoustic classification can help monitor sub-sea animals and detect</li></ul> |

| SDG | How AI Could Help (not exhaustive) |
|---|---|
| Below Water | pollution sources, help remove pollutants, and preserve marine environments to ensure their safety and conservation (see [Andreas et al](#)).<br>• Image classification and satellite imagery analysis can help monitor fish abundance and detect illegal fishing activities, aiding in enforcement and conservation efforts (see [Potineni](#)).<br>• Image-recognition technology on waste facility images and data can help increase recycling rates and reduce plastic pollution in oceans (see [Google](#)). |
| SDG 15: Life on Land | • Satellite imaging can detect deforestation to preserve terrestrial ecosystems<br>• Remote sensing can identify and track endangered species, and detect poaching activities (see [Tuia et al](#)).<br>• Machine learning analytics and audio recognition from rainforests and other ecosystems can help locals derive insights and help root out any ecological threats (see [Google](#)). |
| SDG 16: Peace, Justice and Strong Institutions | • NLP can analyze social media data to identify opinions and emotions on social issues in multilingual communities (see [Mabokela and Schlippe](#)).<br>• Computational statistics and NLP can quantify and reveal patterns of online abuse (see [Tomašev et al](#)).<br>• NLP and other machine learning methods can extract relevant information from legal and judicial documents (e.g., laws, jurisprudence, victim testimonies, and resolutions) and empower human rights advocates (see [Google](#)). |
| SDG 17: Partnerships | • AI-powered platforms can help develop monitoring systems for compliance with SDG targets, analyzing data from multiple sources (see [Sætra](#)).<br>• NLP algorithms can be integrated into collaborative research platforms to analyze and synthesize large volumes of scientific literature, patents, and technical documents (see [Just](#)).<br>• Machine learning analytics on developing country economic indicators and population survey data can help public and private sector entities and governments better target assistance and measure return on investment (see [Google](#)). |

a. SDG 3 - Good Health and Well-being

Ensuring good health and well-being for all remains a global challenge. For example, noncommunicable diseases (NCDs) kill 41 million people every year, accounting for 74% of all deaths globally (see [WHO](#)). In 2020, approximately 800 women died every day from pregnancy or childbirth complications (see [United Nations](#)). The potential of AI to transform healthcare has been recognized by the World Health Organization (see [WHO](#)). Our research found that the largest number of AI use cases (125) fall under SDG3, accounting for 23% of all documented use cases. AI applications offer diverse opportunities, ranging from diagnosis and clinical care to health research and public health interventions.

> **BOX 1**
> ARMMAN uses AI to enhance maternal health outcomes by leveraging a call-based information program, mMitra, to engage pregnant women early on, predict dropout risks and facilitate targeted interventions, ultimately aiming to reduce maternal morbidity and mortality in India. mMitra's AI-powered program also directs support calls.
>
> ARMMAN has seen a 30% increase in engagement among high-risk participants, a 25% increase in the number of pregnant women taking IFA (iron and folic acid) tablets for 90 or more days, and a 47.7% increase in the proportion of women who knew at least three family planning methods. To date, they have reached over 2.9 million women.
>
> *Note: ARMMAN's work in India is supported by Google.org and is a Google research collaborator.*

A few noteworthy AI-powered technologies and initiatives working toward SDG 3: Good Health and Well-being are:

AlphaFold: AI is being used in health research and drug development and is expected to play an important role in improving human understanding of diseases and identifying new disease biomarkers (see Raza). Google DeepMind's AI system AlphaFold solved what is known as the "protein folding problem," by accurately predicting the three-dimensional shape of proteins and other biomolecules like DNA, RNA, ligands and other small molecules, and how they interact together in a cell (see Jumper et al). This breakthrough technology is helping to accelerate drug discovery for cancer, opening up cost-effective opportunities to treat diseases previously neglected (see University Of Toronto), aiding in the development of enzymes to facilitate the breakdown of single-use plastics (see Google), and unlocking insights to grow more resilient crops that require fewer pesticides. The findings are freely available to the scientific community, in partnership with EMBL-EBI, through the AlphaFold Protein Structure database, which has more than 1.8 million users in over 190 countries. AlphaFold has been cited more than 20,000 times in research papers.

Mammography AI: Google partnered with the Japanese Foundation for Cancer Treatment (JFRC), iCAD, Northwestern Medicine, and medical experts, to develop an AI system capable of breast cancer prediction (see Corrado). Early detection not only can lead to better health outcomes but can also ultimately reduce treatment costs. More collective work is needed to improve equitable access to effective cancer treatment once detected.

mMitra: AI algorithms are positively impacting public health outcomes by analyzing large amounts of beneficiary data to provide targeted health promotion and improved identification and response to health outbreaks. ARMMAN leverages AI and a call-based program (mMitra) to improve maternal health outcomes in India by engaging pregnant women, predicting dropout risks, and facilitating interventions. [See Box 1].

b. SDG 4 - Quality Education

Significant challenges persist in achieving SDG 4 (Quality Education). About 59 million children do not attend primary school (see UNESCO). Among those enrolled, 70% of children in LMICs do not have adequate reading and comprehension skills(see World Bank).  If this trend continues, it is estimated that by 2030, only one in six countries will achieve the universal secondary school completion target and approximately 300 million students will lack the basic numeracy and literacy skills necessary for success (see United Nations). In 2020, over 14% of teachers globally lacked minimum qualifications, with Sub-Saharan Africa exhibiting the most pronounced deficiency, where only 60% of pre-primary, 69% of primary, and 61% of secondary teachers met qualification standards, exacerbating disparities in access to quality education (see United Nations).

> **BOX 2**
> Quill utilizes AI to offer immediate personalized feedback and coaching to students in an interactive manner that enhances their reading and writing skills and improves literacy outcomes. The tool is intentionally designed to support teachers by assisting them in identifying and focusing their efforts on students who require the most support.
>
> Quill has helped over 8.9 million students across 35,000 schools improve their writing skills, with almost 2 billion sentences written and support provided to 162,000 teachers.
>
> *Note: Quill.org, based in the US, is a 2019 recipient of support from Google.org.*

A recent report by UNESCO, "AI and Education: Guidance for Policy-Makers," highlights that AI technologies can augment educational experiences while upholding ethical, equitable, and inclusive standards (see UNESCO). Moreover, a growing body of research underscores the potential of AI to revolutionize various facets of education (see [World Economic Forum](#), [UNESCO](#), [Zhang and Aslan](#)). In our research, education-related use cases (38) constitute 7% of documented AI SDG use cases.

Some examples of AI-powered technologies and initiatives focused on SDG 4 (Quality Education) are:

Read Along: Google's speech-based reading tutor app, allows students to learn by reading aloud (see [Google](#)). The app's impact is visible through the data from the pilot study: 64% of participants from the India pilot study with access to the app showed an improvement in reading proficiency, and 95% of parents from the India pilot study said they would continue to use the app even after the pilot study (see [Google](#)). Such innovations help teachers increase student engagement and allow them to provide personalized learning support, particularly in settings where there is a poor teacher-to-student ratio.

Quill: With diverse pedagogies and a need for teacher engagement, Intelligent Tutoring Systems and Dialogue-based Tutoring Systems can support the delivery of personalized learning experiences to supplement teaching. Quill.org uses AI to provide personalized feedback and coaching to students through its AI-driven tool Quill, improving their literacy skills while assisting teachers in identifying and supporting those who need it most, impacting over 8.9 million students. [See Box 2.]

Student Support Tool: AI can also be applied to support learners in their learning journey and assist educational institutions and teachers to streamline administrative processes, facilitate data-driven decision-making, and enhance education outcomes. DataKind — a nonprofit that uses data science and AI to improve the capabilities of social impact organizations — collaborated with City University of New York: John Jay College Criminal Justice, supported by Google.org, to build and sustain a predictive AI Student Support Tool that identifies students most at risk of dropping out of undergraduate studies, helping increase the senior graduation rate from 54% to 86% in just two years through targeted support (see [Google](#)).

c. SDG 13 - Climate Action

Climate change is one of humanity's greatest collective challenges and it requires global collaboration and change to reduce emissions, transition energy systems, and make industries more sustainable. Worldwide, about 3.3 to 3.6 billion people are highly vulnerable to climate change impacts (see [IPCC](#)). In 2023, disasters related to natural hazards affected 93.1 million people (see [CRED](#)), with economic losses amounting to $202.7 billion (see [CRED](#)). AI is considered an effective tool to achieve climate goals (see [Boston Consulting Group](#)). AI can have a transformative effect on climate progress by providing helpful information, predicting climate-related events, and optimizing climate action together with partners (see [Google](#)). Increasingly, AI's role is recognized in both climate change mitigation and adaptation, including the four phases of disaster management: mitigation, preparedness, response, and recovery (see [Sun et al](#)). While AI has the potential to positively influence climate action, it is also important to address the environmental impact associated with it by innovating to increase efficiency and reduce the carbon footprint of workloads (see [Patterson](#)).

Examples of AI-powered technologies and initiatives focused on SDG 13 (Climate Action) include:

Flood Hub: A Google Research initiative on flood forecasting, provides actionable information to help individuals stay safe and allow communities to plan ahead (see [Google](#), [Google](#)). Using satellite imagery, advanced AI, and geospatial analysis, those free real-time forecasts can help protect the most vulnerable communities. The latest advances in AI allow the scaling of flood forecasts to data-scarce and vulnerable areas and narrow the accessibility gap by improving forecasts in regions of Africa and Asia to be similar to those currently available in Europe (see [Nearing et al](#) ). Accurate and more accessible flood forecasts, available up to seven days in advance, are now helping governments and nonprofits in 80 countries, covering 460 million people globally. Work with governments and nonprofits ensures that alerts can reach everyone, regardless of connectivity.

Green Light: A Google Research initiative, uses AI and Google Maps driving trends to model traffic patterns and make recommendations for optimizing existing traffic light patterns (see [Google](#)). By optimizing not just one intersection, but coordinating across several adjacent intersections to create waves of green lights, cities can improve traffic flow and further reduce stop-and-go emissions. Currently live at 70 intersections in 12 cities across four continents, early numbers indicate Green Light's potential for up to a 30% reduction in stops and up to a 10% reduction in emissions at intersections. Recommendations in these cities can also save fuel and lower emissions for up to 30 million car rides per month.

> **BOX 3**
>
> Climate TRACE's emission map platform and its Global Emissions Inventory track emission data through satellite coverage and AI, providing near-real time data for 99.9% of the world's greenhouse gas emissions sources, including tracking the largest 352 million individually emitting sources in the world. Furthermore, its Automated Emissions Reduction solution integrates with smart devices to switch energy use towards renewable energy.
>
> To date, about 11 million tonnes of carbon dioxide emissions have been avoided through 100+ contributing organizations.
>
> *Note: WattTime (a founding member Climate TRACE coalition), based in the US, is a 2019 recipient of support from Google.org.*

The Tree Canopy tool: Part of Google's Environmental Insights Explorer platform, the Tree Canopy tool combines AI and aerial imagery to show where shaded areas are in a city. This helps officials strategically plan tree coverage to mitigate extreme heat and is available in more than 2000 cities globally (see [Google](#)).

Climate TRACE: A nonprofit coalition, leverages satellite data and AI to track 99.9% of global greenhouse gas emissions in near real-time, including 352 million individual sources, while its Automated Emissions Reduction solution has helped avoid 11 million tonnes of CO2 emissions through 100+ organizations. [See Box 3.]

d. AI use cases across the SDGs

While this discussion and the evidence provided above have focused on the application of AI across health, education, and climate, Google is working on AI tools with broad applicability across the SDGs. A few noteworthy AI-powered technologies and initiatives have value across the SDGs:

The 1000 Languages Initiative is enhancing machine translation capabilities for lesser-known languages, helping bridge language barriers, and making global communication more inclusive and accessible. [See Box 4.]

> **BOX 4**
>
> Google researchers are working on the 1000 Languages Initiative, aimed at overcoming the challenge of limited language translation coverage, especially for marginalized communities using lesser-known languages. Google is aiming to expand language translation coverage to the 1000 most spoken languages and bring greater inclusion to billions of people all over the world. The project leverages multilingual models trained with supervised parallel data for over 100 high-resource languages and monolingual datasets for an additional 1000 languages and has already been successful in adding 24 such languages to Google Translate, marking a crucial step towards greater linguistic understanding.

Project Relate is using recognition technology to aid individuals with non-standard speech, providing personalized models for speech recognition, enabling transcription, and facilitating interactions with voice-based assistants for tasks (see [Cattiau](#)).

Google's Data Commons is using AI to bridge the gap between public data availability and usability by organizing and interlinking over 240 billion data points across 450,000 variables from 256 countries. Datasets are sourced from authoritative sources across the globe, including government statistical agencies, international organizations (including UN agencies and the World Bank), and more, and cover a range of topics. Data Commons, an open-source tool, enables policymakers, researchers, and the public to ask questions, visualize and analyze data, and better understand the world's challenges through a natural language search functionality (see [Guha and Manyika](#)).

The discussion above offered evidence from Google's internal and collaborative research, technical applications, and social impact initiatives with work in the field to show AI's potential to help accelerate the SDGs and provide some leading examples of how AI is already helping to move the needle on the SDGs. These examples illustrate just some of Google's technical work and collaborations and Google.org's social impact initiatives that support SDG efforts. The breadth of solutions being developed and deployed indicates an opportunity for collaboration across the ecosystem of actors, addressing specific challenges and creating shared solutions and tools that can further accelerate work toward the SDGs.

## 3. DEPLOYING AI RESPONSIBLY

AI-driven solutions for societal and environmental goals require ethical practices and thoughtful policies to promote development and deployment that are responsible, transparent, and fair, while also maximizing positive impacts for people (see [RELX SDG Resource Centre](#)). Essential practices include transparency and explainability, data privacy and security, fairness and bias mitigation, inclusivity and accessibility, human-centered design, and continuous monitoring and assessment (see [World Economic Forum](#)). This includes policy frameworks — by governments and international organizations — to ensure the positive impact of AI technologies through ethical guidelines, data governance, transparency, inclusivity, accountability, stakeholder engagement, capacity building, monitoring, evaluation, and collaboration.

AI's promise will not automatically translate into progress on the SDGs: policymakers must adopt an enabling policy agenda that empowers the public sector, civil society, and businesses of all sizes to develop and deploy AI to take on specific SDG challenges. An enabling agenda could include: driving cloud-first initiatives to help actors across government, business, and civil society deploy AI systems cost-effectively; supporting workers and organizations who can make progress on SDGs with AI skilling initiatives; modernizing national data systems and supporting trusted data flows to drive the development of high-quality datasets that represent diverse perspectives, languages, and cultures and are relevant to local communities; and pursuing a set of pro-innovation regulatory frameworks that take a risk-based approach to AI governance while enabling the usage of publicly available information (see [Google AI Sprinters](#), [OECD](#)).

Engagement and participation at the local and/or organizational levels are also integral to the success of AI solutions aimed at accelerating progress on the SDGs. Google.org's social impact initiatives with local organizations, deeply rooted in communities, have demonstrated the importance of on-the-ground engagement to drive impactful AI solutions. Engagement and participation across disciplines, geographies, and sectors helps to ensure AI-driven solutions are tailored to community and organizational needs, challenges, and values. Key considerations for advancing engagement and participation include needs assessment, educational programs (awareness-raising and capacity development), cultural sensitivity, consultative decision-making, localization of solutions, and user-friendly interfaces (see [Google](#)).

Building on this understanding of ethical and policy considerations gleaned from case studies and technical applications discussed above, research, expert interviews conducted as part of our collaborative work with partners, and Google's AI principles, we have distilled seven key components that organizations should consider as they develop and deploy AI-driven solutions to advance the SDGs (see [Google AI Principles](#)).

Table 2: Key components for responsibly designing, developing, and deploying AI solutions for SDGs

| Component | Guidance for Effective AI Solutions | Case Example |
| --- | --- | --- |
| Problem suitability for AI | Assess if AI is the right tool for the specific problem. Does the problem have clear inputs and outputs that AI can address? Can AI mitigate risks and maximize benefits? | [1000 Language Initiative](): By employing various evaluation metrics, including human assessments, Google demonstrates the suitability of AI for addressing complex language translation challenges across diverse linguistic contexts. |
| Data access and quality | Secure ethical, high-quality data for the AI solution. Invest in data collection, curation, and validation to ensure accuracy, consent, and reflection of real-world scenarios. | [ARMMAN](): Uses an ML model to deliver life-saving information to pregnant women, prioritizing NGO-led community data stewardship. The tool uses anonymized data and collaborates with public health and field experts to assess socioeconomic indicators comprehensively, ensuring both data quality and ethical sourcing. |
| Stakeholder engagement | Cultivate partnerships with diverse stakeholders — researchers, data providers, governments, community organizations, and end-users — throughout the AI development lifecycle. | [AI in Mammography](): This project actively engages clinicians, patients, health professionals, and universities to inform, design, test, and deploy an AI system enhancing breast cancer detection for radiologists, highlighting the importance of involving a wide range of stakeholders in the AI development process. |
| Accountability systems | Prioritize transparency and responsibility to individuals and communities. Clearly define roles and responsibilities for technical safety and establish mechanisms to address unintended consequences. | [DataKind](): Uses a predictive AI tool to identify students most at risk of dropping out of undergraduate studies. The solution was co-created with feedback from key stakeholders, including academic advisors, administrators, and faculty. They addressed concerns about the ethical use of a predictive model to support student success. |
| Responsible model access and evaluation | Ensure responsible and secure deployment of AI models by investing in expertise, practices, and tools for evaluation, and by engaging with external domain experts. Ensure models are shared in ways that minimize risk and consider long-term societal implications. | [AlphaFold:]() Google DeepMind consulted over 50 experts across relevant domains such as biosecurity, research and industry to analyze the capabilities and potential risks of each AlphaFold model. These analyses were an integral part of each model release decision and the mode of release. |
| User-centered design | Co-create inclusive solutions with diverse user groups, including frontline workers, end-users, community members, peer organizations, and those with limited access or who speak less common languages. | [Project Relate](): Employs personalized AI speech recognition to facilitate communication among individuals with non-standard speech. It leverages over a million speech samples and allows users to train the algorithm to understand their unique speech patterns, making it truly user-centric. |
| Scalability and sustainability | Plan for scale and sustainability from the start. Test and refine AI solutions in real-world scenarios, while continuously assessing their | [Quill.org](): This AI-powered tool provides feedback to improve writing skills. It was designed with teachers and scalability in mind and has successfully reached more than 8.9 |

| | long-term impacts. | million students, showcasing the potential of AI for widespread impact. |
|---|---|---|
| Continuous experimentation and learning | Foster a culture of collaboration and learning by building an organizational culture and systems that encourage and facilitate ongoing experimentation, measurement, evaluation, and adaptation of AI solutions. | [Flood Hub](): This project leverages machine learning for flood forecasting. By continuously refining its models, integrating new data sources, and collaborating with local governments and communities, Flood Hub enhances the accuracy and accessibility of its predictions, demonstrating the importance of ongoing learning and adaptation in AI systems. |

4. CONCLUSION

AI holds significant potential to help address some of humanity's most pressing challenges, and as seen in the examples above, it is already helping accelerate progress towards the SDGs. As we approach 2030, the importance of AI is expected to grow. AI can help reimagine existing solutions and pioneer new pathways for innovation and impact: from enhancing healthcare delivery and personalizing education to advancing climate action. This brief offered evidence from Google's internal and collaborative research, technical applications, and social impact initiatives to show AI's potential to help accelerate the SDGs and provided some leading examples with current and future impact. Advances in AI capabilities, including generative AI, machine learning, and robotics, are expanding the repository of use cases across the SDGs. Delivering on and scaling AI's potential for impact on the SDGs is a collaborative endeavor that requires work across companies, universities, NGOs, governments, and individuals to have the real-world impact we seek.

It is also important to acknowledge that AI is not a solution for every challenge or opportunity, but one part of a set of solutions. Successful AI deployment depends on several key enablers, including data, compute, and technical skills. The effectiveness of AI tools also requires grappling with responsible AI considerations, including ethical principles, risks, and data privacy. This emphasizes the need for co-creation to ensure AI-driven solutions are sensitive to and effective in addressing diverse social contexts and challenges. Ultimately, partnerships between government, academia, industry, and civil society are critical to leveraging the potential of AI to address the most pressing societal challenges and collectively benefit society.

Acknowledgements

Many thanks to the many individuals at Google who have contributed to the AI research, initiatives, innovations, and social impact work we cite in this research brief, including: Abhishek Bapna, Adi Gerzi Rosenthal, Adi Mano, Aisha Walcott-Bryant, Akshat Sharma, Alexander Pritzel, Aliza Hoffman, Alon Harris, Amy Chung-Yu Chou, Agata Laydon, Andrew Cowie, Andrew Ballard, Andrew Senior, Anna Carter, Anna Potapenko, Anna Sard, Anton Kast, Aparna Taneja, Arink Verma, Augustin Žídek, Avinatan Hassidim, Avishai Zagoury, Ayelet Benjamini, Bernardino Romera-Paredes, Biswajeet Mallik, Bob MacDonald, Chris Kelley, Clemens Meyer, Dan Golden, Dan Karliner, Dan Morris, Dan Tse, Dana Cohen, Danny Veikherman, David Reiman, David Silver, Deborah Cohen, Dem Gerolemou, Demis Hassabis, Dotan Emanuel, Ela Gruzewska, Eliav Buchnik, Fadi Biadsy, Fereshteh Mahvar, Françoise Beaufays, Gal Weiss, Greg Corrado, Grey Nearing, Hadar Shemtov, Hadas Fester, Irwin Boutboul, Jack Haddad, Jeff Dean, Jennifer Holland, Joel Conkling, Johan Schalkwyk, John Jumper, Jonas Adler, Josh Abramson, Julie Cattiau, Juliet Rothenberg, Kasumi Widner, Katherine Chou, Kathrin Probst, Kathryn Tunyasuvunakool, Kevin Holst, Komal Singh, Koray Kavukcuoglu, Liat Ben Rafael, Lihong Xi, Lin Yang, Luke Barrington, Luke Leonhard, Mansi Kansal, Marc Wilson, Matheus Vervloet, Megumi Morigami, Melvin Johnson, Michael Figurnov, Michal Zielinski, Milind Tambe, Moriah Royz, Nitzan Tur, Olaf Ronneberger, Omer Litov, Oren Gilon, Ori Rottenstreich, Oriol Vinyals, Pan-Pan Jiang, Perry Nelson, Philip Nelson, Pushmeet Kohli, Rebecca Moore, Richard Evans, Rishub Jain, Ronit Levavi Morad, Rus Heywood, Ruth Alcantara, Sebastian Bodenstein, Shai Ferster, Shantanu Sinha, Shawn Xu, Shravya Shetty, Stanislav Nikolov, Stig Petersen, Sunil Muthiyan, Sunny Jansen, Tiffany Chen, Tiffany Deng, Tim Green, Tina Bellon, Tom Kalvari, Uche Okonkwo, Yaniv Elimor, Yazan Safadi, Yoli Spigland, and Yuval Carny.

We are also grateful for the Google colleagues who contributed to the ideas and insights put forward in this brief, including Alex Diaz, Andre Nakazawa, Andrew Kim, Anna Koivuniemi, Anoop Sinha, Antonia Gawel, Brian Juhyuk Lee, Dan Altman, Dorothy Chou, Erin Hattersley, Guy Ben-Ishai, Hannah Bill, JamieSue Goodman, Jen Carter, Joel Straker, Joseph Hassine, Karla Palmer, Kevin Brege, Marsden Hanna, Melike Yetken Krilla, Michael Pisa, Nicholas Bramble, Prem Ramaswami, Ricki Meyer, RV Guha, Stuart McLaughlin, and Yael Maguire.

Collaborative work with hundreds of nonprofits and other social impact organizations have not only yielded tremendous impact but have also been central to the insights and cases presented in this brief. While we are unable to name all partners individually, we extend our thanks for their partnership and shared commitment to driving positive change through AI. Continued research and attention will push forward the collective efforts needed to best deliver on AI's potential for social impact and to accelerate progress on the SDGs. We appreciate the extensive body of research conducted by various organizations and individuals, which has informed and enriched this brief, including the published work we cite.

The views expressed herein are those of the authors and do not necessarily reflect the views of Google or Alphabet, or those we have been in dialogue with.

https://www.mckinsey.com/~/media/mckinsey/business%20functions/quantumblack/our%20insights/the%20state%20of%20ai%20in%202022%20and%20a%20half%20decade%20in%20review/the-state-of-ai-in-2022-and-a-half-decade-in-review.pdf

Mirestea, Alin, et al. *Powering the Digital Economy: Opportunities and Risks of Artificial Intelligence in Finance*. IMF e-Library. October 2021.
https://doi.org/10.5089/9781589063952.087

Mitchell, Colin, et al. *Artificial Intelligence for Genomic Medicine*. PHG Foundation, May 2020.
https://www.phgfoundation.org/publications/reports/artificial-intelligence-for-genomic-medicine/

Murphy, Robin. *Search and Rescue Robotics*. Springer Handbook of Robotics, 2nd Edition, 1151-1173, 2014.
https://link.springer.com/referenceworkentry/10.1007/978-3-540-30301-5_51

Nearing, Grey, et al. *Global prediction of extreme floods in ungauged watersheds*. Nature, 2024.
https://doi.org/10.1038/s41586-024-07145-1

Nelson, Perry and Aisha Walcott-Bryant. *6 Ways Google Is Working with AI in Africa*. Google Africa Blog, June 2023.
https://blog.google/intl/en-africa/company-news/6-ways-google-is-working-with-ai-in-africa/

*New Study Uses AlphaFold and AI to Accelerate Design of Novel Drug for Liver Cancer*. University of Toronto, January 2023.
https://acceleration.utoronto.ca/news/new-study-uses-alphafold-and-ai-to-accelerate-design-of-novel-drug-for-liver-cancer

Onyeaka, Helen. *Using Artificial Intelligence to Tackle Food Waste and Enhance the Circular Economy: Maximizing Resource Efficiency and Minimising Environmental Impact: A Review.* MDPI, July 2023.
https://www.mdpi.com/2071-1050/15/13/10482

Patterson, David. *Good News About the Carbon Footprint of Machine Learning Training*. Google Research, February 2022.
https://research.google/blog/good-news-about-the-carbon-footprint-of-machine-learning-training/

Potineni, Abhinav. *Edge Technology Based Artificial Intelligence System for Ocean Patrol and Surveillance*. International Journal of Scientific Research in Computer Science, Engineering and Information Technology, 182-191. July 2022.
https://ijsrcseit.com/home/issue/view/article.php?id=CSEIT22845

*Products and Services - AI/ML Training and Technology.* Karya.in, 2024.
https://karya.in/ai-ml-technology/

Qiu, Shuang, et al. *Systematic Review on Wearable Lower Extremity Robotic Exoskeletons for Assisted Locomotion.* Journal of Bionic Engineering. 2023.
https://typeset.io/pdf/systematic-review-on-wearable-lower-extremity-robotic-3fsjrwq5.pdf

Rana, Meghavi and Meghan Bhusan. *Machine learning and deep learning approach for medical image analysis: diagnosis to detection.* Springer, 2023.
https://link.springer.com/article/10.1007/s11042-022-14305-w

*Read Along - An Android app from Google*.
https://readalong.google/impact/

RELX SDG Resource Centre. *Artificial Intelligence.* RELX SDG Resource Centre, 2024.
https://sdgresources.relx.com/artificial-intelligence-0#:~:text=However%2C%20the%20integration%20of%20AI,concerns%2C%20and%20the%20digital%20divide

*Remote Sensing Technology Market.* Precedence Research, April 2023.
https://www.precedenceresearch.com/remote-sensing-technology-market#:~:text=The%20global%20remote%20sensing%20technology,11.79%25%20from%202023%20to%202032
18